\documentclass[pra,aps,twocolumn,amsmath,amssymb,floatfix,reprint,footinbib,showpacs,superscriptaddress]{revtex4-2}

\usepackage[dvipsnames]{xcolor}
\usepackage{amsmath}
\usepackage{amssymb}
\usepackage{amsfonts}
\usepackage{tikz}
\usepackage{dsfont}
\usepackage{url}
\usepackage{graphicx}
\usepackage[colorlinks]{hyperref}
\hypersetup{%
	plainpages=true,
	breaklinks=true,
	hypertexnames=false,
	pageanchor=true,
	colorlinks=true,
	linkcolor={blue},
	citecolor={red},
	urlcolor={blue},
	anchorcolor={black}
}
\usepackage{bbm}
\usepackage[draft,inline,nomargin,index,marginclue]{fixme}
\fxsetup{status=draft}
\graphicspath{{.}{./figs/}}
\newcommand{\ket}[1]{{\vert #1 \rangle}}

\newcommand* {\bra}[1]{\ensuremath{\langle {#1} |}}
\newcommand{\imag}{\textrm{i}}

\newcommand{\ave}[1]{{\langle #1\rangle}}

\begin{document}
\title{Ultrastrong waveguide QED with giant atoms}
\author{Sergi Terradas-Briansó}
\author{Carlos A. Gonz\'alez-Guti\'errez}
\affiliation{Instituto de Nanociencia y Materiales de Arag\'on (INMA), CSIC-Universidad de Zaragoza, Zaragoza 50009, Spain}
\author{Franco Nori}
\affiliation{Department of Physics, The University of Michigan, Ann Arbor, Michigan 48109-1040, USA}
\affiliation{Theoretical Quantum Physics Laboratory, RIKEN Cluster for Pioneering Research, Wako-shi, Saitama 351-0198, Japan}
\affiliation{RIKEN Center for Quantum Computing (RQC), 2-1 Hirosawa, Wako-shi, Saitama 351-0198, Japan}
\author{Luis Martín-Moreno}
\author{David Zueco}
\affiliation{Instituto de Nanociencia y Materiales de Arag\'on (INMA), CSIC-Universidad de Zaragoza, Zaragoza 50009, Spain}
\date{\today}
\begin{abstract}
Quantum optics with giant emitters has shown a new route for the observation and manipulation of non-Markovian properties in waveguide-QED. In this paper we extend the theory of giant atoms, hitherto restricted to the perturbative light-matter regime, to deal with the ultrastrong coupling regime. Using static and dynamical polaron methods we address the low energy subspace of a giant atom coupled to an Ohmic waveguide beyond the standard rotating wave approximation. We analyze the equilibrium properties of the system by computing the atomic frequency renormalization as a function of the coupling characterizing the localization-delocalization quantum phase transition for a giant atom. We show that virtual photons dressing the ground state are non-exponentially localized around the contact points but decay as a power-law. Dynamics of an initially excited giant atom are studied, pointing out the effects of ultrastrong coupling
on the Lamb shift and the spontaneous emission decay rate. Finally we comment on the existence of the so-called oscillating bound states beyond the rotating wave approximation.
\end{abstract}

\maketitle

\section{Introduction}%

The coupling of a single quantum emitter to a continuum of electromagnetic modes is an important problem since the birth of quantum theory \cite{Weisskopf1930}. 
Current experiments, involving different technological platforms have shown that propagating photons can be coupled efficiently to localized quantum emitters. This field, known as waveguide quantum electrodynamics (wQED), has received a lot of attention due to the interesting theoretical and experimental applications \cite{Wilson2017,Sasha2021,Tao2018}.
In most scenarios, emitters are described as point-like particles of negligible size compared with the wavelength of the electromagnetic radiation. This justifies the standard dipole approximation widely employed in quantum optics.
In recent years, however, experiments with artificial emitters have led to a reconsideration of atoms as point-like matter \cite{GustafssonSci2014,AnderssonNatPhys2019}. 
In the literature, they are called \emph{giant atoms}, for obvious reasons. 
As a consequence of the non-local light-matter interaction, new phenomena have been reported. Examples are non-Markovian emission \cite{KockumPRA2017,AnderssonNatPhys2019, LonghiOpt2020}, tunable decay rates and Lamb shifts \cite{KockumPRA2014,KockumPRL2018,KannanNat2020}, engineering of energy levels \cite{VadirajPRA2021}, as well as bound states emerging from interference between coupling points, including oscillating \cite{KockumPRR2020, GuoPRA2020} and chiral \cite{WangPRL2021} bound states. 
In addition, bound states originating from photonic band edges for giant atoms have been studied in \cite{ZhaoPRA2020}.
The large size of the system also allows for a giant emitter to be coupled to a waveguide in between the connection points of other giant atoms. The many possible configurations can lead to decoherence-free interactions between giant emitters \cite{KockumPRL2018, KannanNat2020} or nonreciprocal excitation transfer \cite{DuPRA2021}.
See Ref. \cite{KockumReview2021} for a recent overview of the field.

The breakdown of the dipolar approximation leads to the appearance of deviations form Markovian dynamics.
These typically arise from the coupling of quantum emitters to structured environments with non-flat spectral functions \cite{vats1998,breuer, deVega2017}. 
However, it has been shown that retardation effects can induce strong non-Markovian features whenever coherent feedback is allowed to influence the dynamics \cite{Dorner2002, Pichler2016, Tuffarelli2013, Kanu2020, DistantEmitters2021, Regidor2020, Regidor2021, Wen2019}.
Giant emitters fall naturally into this last category of non-Markovian systems \cite{KockumReview2021} and they have been a relevant topic in waveguide QED systems.

Another assumption that is being reconsidered, thanks to experiments, is the fact that photons are weakly coupled to matter, so their interaction can be described in a perturbative way.
Several experiments have reached the so-called ultrastrong coupling regime (USC) between light and single quantum
emitters, both in cavity \cite{Niemczyk2010,forn2010,Yoshihara2017} and waveguide QED \cite{forn2017,martinez2019,leger2019}. 
In the USC regime higher order processes,
than the creation (annihilation) of one photon by annihilating (creating) one matter excitation play a role. 
Then, the rotating wave approximation (RWA) for the interaction breaks
down, the atomic bare parameters get renormalized, and the ground state becomes nontrivial.
This has interesting consequences. Some of them are the possibility of transforming virtual onto real photons by perturbing the ground state \cite{Ciuti2005, Stassi2013, QiKai2018, gheeraert2018, SanchezBurillo2018, Cirio2017}, the localization-delocalization transition \cite{peropadre2013,shi2018}, or the possibility to perform non-linear optics at the single and zero photon limit \cite{sanchez2014,sanchez2015,Stassi2017,Kockum2017, Chen2017, Kockum2017b}. 
Reviews for light-matter interactions in the USC regime can be found in \cite{KockumUSCReview2019,forn2019}.

In this work, we discuss the low energy physics (both at and out of equilibrium) of a giant atom coupled to a continuum in the USC regime.
To do so, the light-matter coupling is treated within the spin-boson model. 
In the USC this is 
a paradigmatic example of a non analytically solvable model \cite{Weiss}. 
Different techniques are available in the literature to deal with it, such as Matrix-product
states \cite{peropadre2013,sanchez2014,sanchez2015}, density matrix renormalization
group \cite{prior2010}, hierarchical equations of motion or pseudomodes methods \cite{Lambert2019}, and path integral \cite{grifoni1998,lehur2010},
 polaron-like \cite{shi2018,sanchez2019,silbey1984,Bera2014,DiazCamacho2016,Zueco2019,ashida2021} or Gaussian approaches \cite{shi2018b}. 
 During the completion of this work, it has been recently reported \cite{noachtar2022nonperturbative} how to use matrix product states to describe the dynamics of giant atoms in a waveguide in the USC regime. 
 
 In this paper we employ polaron-like techniques, complementing and extending their work.
We examine the renormalization of atomic parameters and provide expressions for them. 
We prove the the existence of the localization-delocalization transition in giant emitters, as well as a profile of the virtual photons in the ground state which we characterize for both phases. 
Regarding the dynamics, we discuss the spontaneous emission, its rate and the Lamb shift in the USC regime. 
Within the non-Markovian regime we provide numerical results and analytical expressions for the emission and the existence of bound states, also oscillating ones.

The rest of the manuscript is organized as follows. In Sec. \ref{sec:Model} we introduce the theoretical model, including the discrete model for the waveguide as well as the spin-boson Hamiltonian and spectral density of the system. In Sec. \ref{sec:Polaron} we describe the polaron formalism and apply it to our model to reach the effective Hamiltonian used throughout the work. In Sec. \ref{sec:Equilibrium} we analyze the equilibrium properties of the system, including its ground state and renormalization of the transition energy leading to the discussion of the quantum phase transition. In Sec. \ref{sec:DecayRateLambShift} we compute the Lamb-Shift and effective decay rate for the system. In Sec. \ref{sec:nonMarkDyn} we study different cases of the non-Markovian dynamics of the system, using numerical simulations and approximate analytical expressions with a special focus on oscillating bound states. Finally, a summary and conclusions of this work are given in Sec. \ref{sec:Conclusion}.
\section{Spin-Boson model for a giant emitter}\label{sec:Model}%
\begin{figure}
    \centering
    \includegraphics[width = \columnwidth]{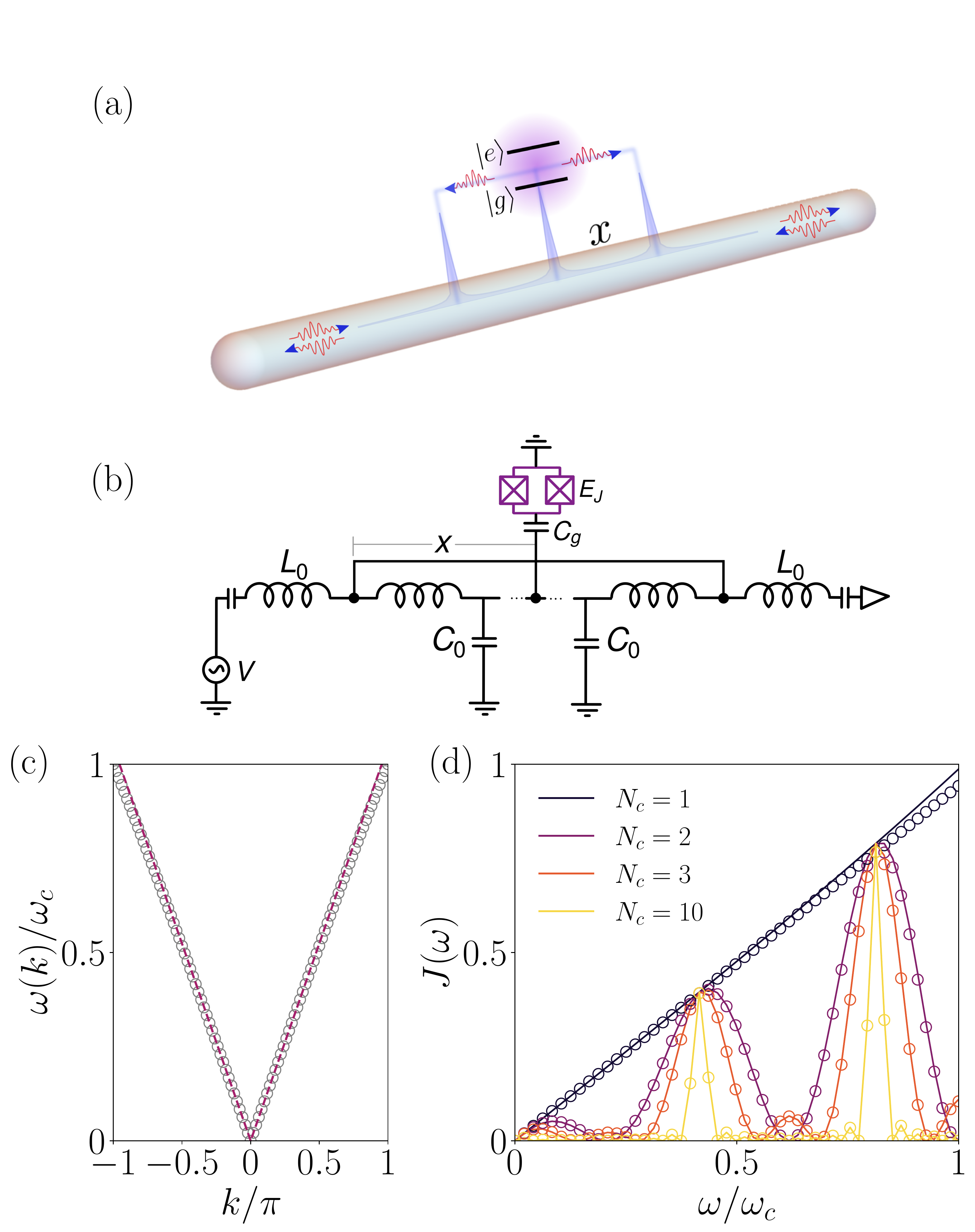}
    \caption{(a) Pictorial illustration of the giant emitter with three connection points. (b) Schematic of a circuit-QED implementation of a giant atom coupled to an Ohmic waveguide with three connection points. 
    (c) Dispersion relations for the discrete and continuous models for the Ohmic waveguide. The group velocity for the waveguide is $v_g = c = 1$ throughout this work.
    (d) Spectral function $J(\omega)$ for a continuous dispersion relation 
  $\omega_k = v_g \lvert k \rvert$ 
  in solid lines and, in circles, the corresponding for the discrete model from Eq. \eqref{disp::rel} with spacing between coupling points $x = 5 \delta x$ and coupling strength $\alpha = 0.1$.
   The cutoff frequency is $\omega_c = 3$ and the number of modes is $N = 300$ for both plots (c) and (d).}
    \label{fig:fig1}
\end{figure}

Let us consider a giant atom well described by a two-level system or qubit with energy splitting $\Delta$.  Furthermore, we assume that the coupling of such qubit to a one-dimensional waveguide is through a discrete set of \emph{contact points}, see Fig. \ref{fig:fig1}(a).
This covers the current realizations with superconducting qubits as reported in Ref. \cite{KockumPRL2018, KannanNat2020}.
So the (giant) atom and waveguide can be described by the following spin boson model:
\begin{eqnarray}
\label{spin::boson}
H=\frac{\Delta}{2}\sigma^{z}+\sum_{k}\omega_k a^{\dagger}_{k}a_{k}+\sigma^x\sum_{k}\left(\tilde g_{k}a_{k}+\rm {h.c.}\right),
\end{eqnarray}

The waveguide modes are found by diagonalizing the corresponding microscopic model via the standard procedure described in \cite{DevoretFluctuations}. A discretization length is chosen $\delta x = L/N$, being $L$ the length of the transmission line and $N$ the number of propagating modes.
For a linear medium in one dimension, discretization yields the LC-chain, with set of momenta $k_n=2\pi n/L,$  $n\in\{-N/2, ..., N/2\}$ and the dispersion relation to use in \eqref{spin::boson}:
\begin{eqnarray}
\label{disp::rel}
\omega_k =\omega_c\sqrt{2-2\cos \left(k_{n}\delta x\right)} \; .
\end{eqnarray}
Here, $\omega_c=v_{g}/\delta x$ is the cutoff frequency.

The emitter-field couplings \cite{KockumReview2021} are dependent on the positions of the coupling points as:
\begin{eqnarray}
\label{gk}
\tilde g_{k} =\frac{g_{k}}{N_c}\sum_{j=1}^{N_c}e^{i k x_{j}}.
\end{eqnarray}
Here, $g_k$ is the coupling strength of a single contact point to the mode $k$, which we assume to be equal at every connection, and $N_c$ the total number of contact points at positions $x_{j}$. 
$1/N_c$ 
is a normalization factor chosen to ensure that in the zero distance limit $\tilde g_k = g_k$, that is, it fixes the total   coupling strength and facilitates the comparison for different $N_c$. Besides, it has a well defined limit as $N_c\to \infty$. 

Hamiltonian \eqref{spin::boson} is completed with the couplings that can be written as
\begin{equation}
\label{eq:gk}
g_k=g\sqrt{\frac{\omega_k}{2 L}} \; ,    
\end{equation}
where $g = \sqrt{\pi v_g \alpha}$ and $\alpha$ denoting the coupling strength to each contact point.

Spin-boson models are characterized by their spectral function:
\begin{equation}
    J(\omega)\equiv 2\pi\sum_{k}|\tilde{g}_{k}|^2\delta(\omega-\omega_k).
\end{equation}
The spectral function encapsulates all the information about the bath frequency modes and their coupling to the two-level system \cite{weiss2012quantum}. 
The discretization, just sketched, guarantees that in the continuum limit $N\to\infty$ ($\delta x\to0$), 
$\omega_k\approx v_{g}|k|$, see Fig. \ref{fig:fig1}(c), and
\begin{eqnarray}
\label{J::omega}
J(\omega)= J_{\rm {Ohm}}(\omega)G(\omega) \; .
\end{eqnarray}
The Ohmic part, $J_{\rm Ohm}(\omega)=
\pi\alpha\omega $, 
comes from the local coupling to a one dimensional continuum, while the modulation function
\begin{equation}
 G(\omega)=\frac{1}{N_c ^2} \; \sum_{j,l}e^{i\omega(x_j-x_l)/v_{g}}   
\end{equation}
arises from interference caused by the multiple coupling points.
For equidistant contact points with distance $x$, the modulation function simplifies to
\begin{equation}
\label{eq:Gw}
 G(\omega)=\frac{1}{N^{2}_{c}}
 \frac{1-\cos (N_{c}\omega x/v_{g})}{1-\cos(\omega x/v_{g})}.
 \end{equation}
 Figure \ref{fig:fig1}(d) shows the spectral function of the waveguide and its modification for different $N_c$, compared to the small emitter limit $N_c=1$ for both the discrete (open circles) and continuous descriptions (solid lines).
 The inter distance $x$ is fixed, so the main peaks coincide for all $N_c$. 
 On the other hand, as the contact points increase, the peaks become narrower with a width $\propto N_{c}^{-1}$. 

\section{Effective RWA models in the USC: Polaron theory for the giant atom} \label{sec:Polaron}%

The low-energy spectrum of
a spin-boson model \eqref{spin::boson} can be well approximated by an
effective excitation-number-conserving Hamiltonian derived
from a polaron transformation \cite{Bera2014, DiazCamacho2016, Zueco2019}. 
Furthermore, it has been shown to be accurate for various realizations of the model,
e.g. considering multiple emitters \cite{JRR2020, DistantEmitters2021} and for different functional forms of the spectral function \cite{Zueco2019}.
In this section we summarize the main aspects of the static and dynamical polaron theory in order to proceed with its application to the case at hand, a giant atom beyond the rotating wave approximation.
The unitary transformation 
\begin{equation}
\label{Up}
 U_p=\exp\left[-\sigma^{x}\sum_{k}\left(f_{k}a^{\dagger}_{k}-f^{*}_{k}a_{k}\right)\right]   
\end{equation}
disentangles the atom and waveguide,
by choosing the displacements $f_k$ such that the ground state of $H_{p}=U^{\dagger}_{p}HU_{p}$ to be as close
as possible to $| g \rangle \otimes |{\bf 0 } \rangle$ the ground state of the uncoupled atom
$|g\rangle$ and waveguide $|{\bf 0} \rangle$.
For \eqref{spin::boson} this is equivalent to minimize the ground state energy ${\rm min}_{f_k} \{ \langle {\bf 0} | \langle g | \, U_p^\dagger H U_p \, | g \rangle|{\bf 0 } \rangle \} $.
It turns out that [Cf. Eq. \eqref{gk}],
\begin{equation}
    \label{eq:fks}
    f_{k}=\frac{\tilde{g}_{k}} {\omega_k+\Delta_r }, 
\end{equation}
with,
\begin{equation}
\label{deltar}
   \Delta_r=\Delta\exp\left(-2\sum_{k}|f_{k}|^2\right).
\end{equation}
Both, $\Delta_r$ and $f_k$ are related by a self-consistent equation that can be solved numerically. Once such parameters are found, we can obtain all the properties of the model. 

Within the scope of the present work, we can restrict our treatment to the low-energy sector, where the polaron model can be well approximated by the effective number-preserving Hamiltonian,
\begin{eqnarray}
\label{Hp}
H_{p}&\approx& H_{\rm eff}
\nonumber\\ &=& 
\frac{\Delta_r}{2}\sigma^{z}+\sum_{k}\omega_k a^{\dagger}_{k}a_{k}
+2\Delta_{r}\sum_{k}f_k\left(\sigma^{+}a_{k}+\rm{h.c.}\right) \nonumber\\
&+&V_{\rm local}+E_{\rm ZP}, 
\end{eqnarray}
where $V_{\rm local}=-2\Delta_r\sigma^{z}\sum_{k,k^{\prime}}f_{k}f_{k^{\prime}}a^{\dagger}_{k}a_{k^\prime}$, and
\begin{equation}
\label{eq:ZPEn}
E_{\rm ZP}=-\Delta_r/2+\sum_{k}f_{k}\left[\omega_{k} f_{k}-\tilde{g}_{k}-\tilde{g}^{*}_{k}\right].    
\end{equation}

%


\subsection{Lab and polaron frames}

Hamiltonian \eqref{Hp} is rather convenient for calculations, because it commutes with the excitation operator $N_e =\sigma^{+}\sigma^{-}+\sum_{k}a^{\dagger}_{k}a_{k}$. This allows the use of the standard methods for the study of waveguide-QED within the RWA. In the polaron frame the ground state is trivial and the dynamics split in subspaces of different number of excitations. 
Expected values of observables in the polaron frame are of minor physical relevance but they are convenient for calculations because the physical observables can be found in terms of them.
Since measurements are performed in the Lab frame, where Hamiltonian \eqref{spin::boson} is expressed, it is mandatory to find the relation between both pictures. 
In what follows, observables with superscript $^p$ are observables computed in the polaron frame, i.e.
\begin{equation}
    O^p := \langle \psi^p(t) |   O  | \psi^p(t) \rangle = 
    \langle \psi(t) U_p^\dagger |  O | U_p  \psi(t) \rangle  \; ,
\end{equation}
whereas actual observables are given by
\begin{equation}
    O = \langle \psi(t) |   O  | \psi(t) \rangle  = \langle \psi^p(t) |  U_p  O U_p^\dagger | \psi^p(t) \rangle \; .
\end{equation}
With this, the atomic excitation probability can be written as, 
\begin{align}
\nonumber
P_e =& \frac{\Delta_r}{\Delta}\left[P^p_e+2\Re\left\{c\sum_k f_k \phi_k^{*}\right\}+2\sum_{k k^\prime }f_{k}f^{*}_{k^{\prime}}\phi^{*}_{k}\phi_{k^\prime}\right]\\
&+P^{\rm GS}_e\;,
\label{eq:PolLabExcit}
\end{align}
where $c = \bra{\bf 0} \otimes \bra{g} \sigma^{-}\ket{\psi^p}$ and $\phi_k = \bra{\bf 0}\otimes\bra{g} a_k \ket{\psi^p}$ are the amplitudes for the excited state and the $k$-mode field of an arbitrary state in the polaron frame, respectively. 
The first and last terms of Eq. \eqref{eq:PolLabExcit} are the probability of excitation in the polaron frame and in the ground state, respectively. 
Concretely,
\begin{equation}
    \label{eq:GSExcit}
    P^{\rm GS}_e = \frac{1}{2}\left(1+\ave{\sigma^{z}}_{\rm{GS}}\right) = \frac{1}{2}\left( 1 - \frac{\Delta_r}{\Delta}\right)
\end{equation}
We also note that to return to the laboratory framework, both the atomic and field amplitudes are needed. This is a consequence of the \emph{non-local} character of $U_p$ in equation \eqref{Up}, 
which mixes
matter and light operators.

Last but not least, we will be interested in the temporal evolution of the occupation of mode $n_k$. In terms of quantities in the polaron frame we obtain the relationship:
\begin{align}
n_k(t) =
 n^{\rm GS}_k +  | \phi_k(t) |^2 -2 \Re\left[c(t)\phi_k(t)f_k\right]  \; .
 \label{eq:PolLabField}
\end{align}
The same comments as for $P_e$ can be repeated here. Both relations will be used through this work. 

\section{Equilibrium properties}\label{sec:Equilibrium}%
For sufficiently weak atom-waveguide coupling, the ground state is well approximated by the trivial vacuum $|g\rangle \otimes|{\bf 0} \rangle$. 
This is consistent with performing the RWA on \eqref{spin::boson}.
A first consequence of entering the USC regime
is that strong light-matter correlations are formed. This is easily understood 
with the polaron \emph{ansatz}, since the actual ground state (GS) of \eqref{spin::boson} can be approximated by
\begin{align}
\label{gs}
\nonumber
\ket{\psi_{\rm GS}} 
& \cong
U_p \, |g\rangle\otimes |{\bf 0} \rangle 
\\
& = 
\frac{1}{\sqrt{2}}\left(\ket{+}\prod_{k}\ket{-f_k}-\ket{-}\prod_{k}\ket{f_k}\right),
\end{align}
where $\ket{f_{k}}=D(f_{k})\ket{0_{k}}$ is a $k$-mode coherent state, being $D(f_{k})=\exp\left({f_{k}a^{\dagger}_{k}-f^{*}_{k}a_{k}}\right)$ the bosonic displacement operator.
States, $| \pm \rangle = \frac{1}{\sqrt{2}} ( |g \rangle \pm |e \rangle)$, are the atom symmetric (antisymmetric) superpositions.
The state in Eq. \eqref{gs} is a multiphoton Schr\"odinger cat state.
Its photon number can be obtained via $\bra{\psi_{ \rm GS}}a^{\dagger}_{k}a_{k}\ket{\psi_{\rm GS}}=|f_{k}|^{2}$. 
The photonic profile in position space can be recovered via a discrete Fourier transform, 
\begin{equation}
    \label{eq:FourierTransf}
    f_{x} =  \frac{1}{N_c}\sum_{j,k}f_{k}e^{ik(x-x_{j})},
\end{equation}
which indicates that the photonic amplitudes are superpositions of small emitter contributions $f_{k}$ centered at each coupling point to the waveguide. 

The ground state, together with the atomic renormalization frequency $\Delta_r$ in Eq. \eqref{deltar}, encapsulate the equilibrium properties at zero temperature. In particular, the existence of virtual excitations, both in the atom and in the photonic field, as well as the existence (or not) of a quantum phase transition. 
It is known that the spin boson model undergoes a localization-delocalization transition when $\Delta_r \to 0$ \cite{Leggett1987}.
Again, this transition can also be understood within the polaron formalism. If we look at $\eqref{Hp}$, when $\Delta_r =0$, the ground state is degenerate, so the gap closes and a quantum phase transition can occur.

\begin{figure}
    \centering\includegraphics[width=\columnwidth]{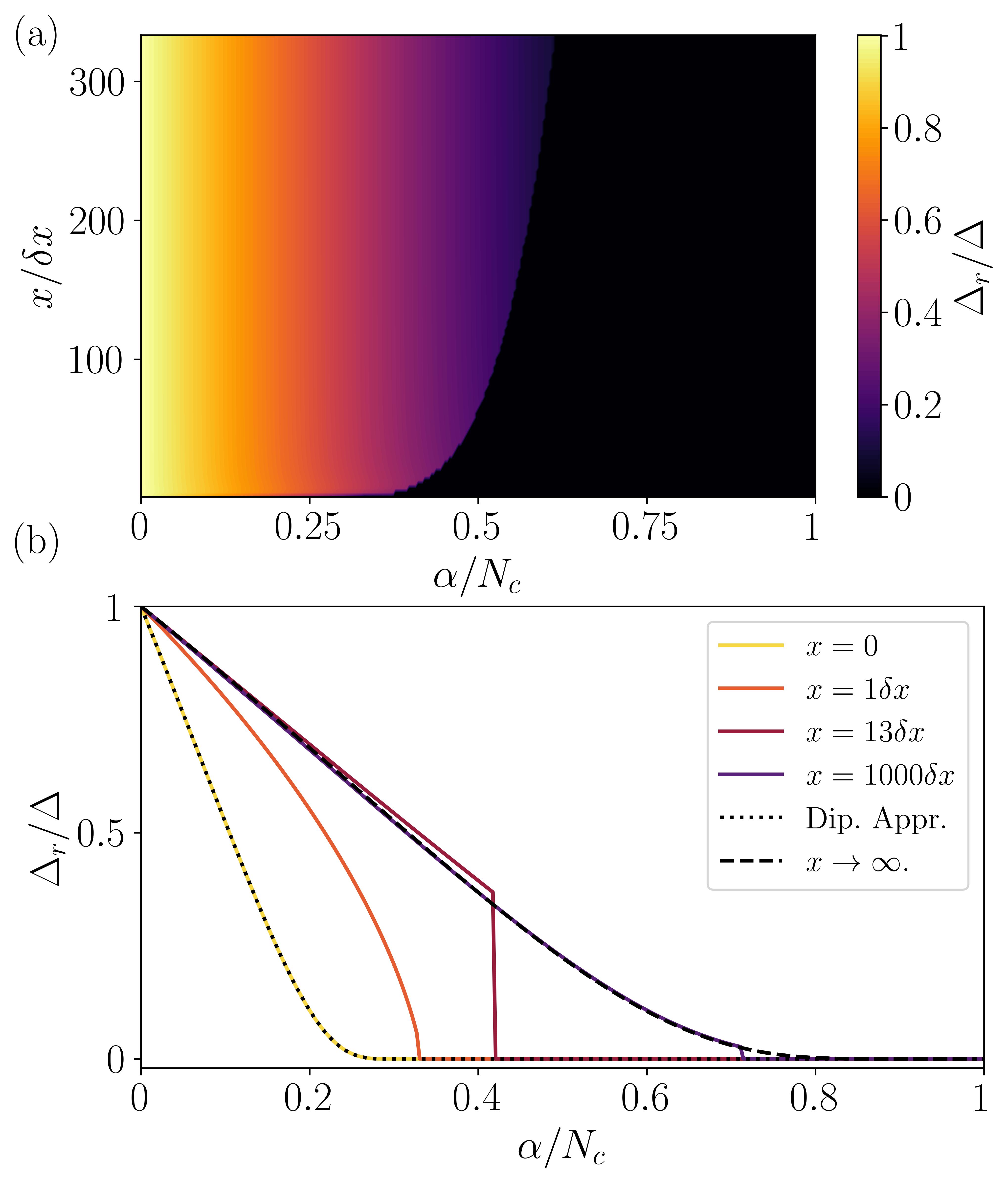}
\caption{(a) Renormalization of the two-level system energy with the coupling strength $\alpha$ and the distance between coupling points $x$ for $N_c = 3$. (b) A set of specific values for the distance $x$ between connections showing the phase transition as $\alpha$ increases. For limiting cases we have analytical expressions, for intermediate distances, the transition is more abrupt.} 
    \label{fig:DeltarPanel}
\end{figure}

\subsection{Atom excitations, renormalization and the existence of a QPT}

A consequence of light-matter entanglement in the ground state is that the atom is dressed by the quantum fluctuations of waveguide photons. This is reflected in a renormalization of the dressed atomic frequency, see $\Delta_r$ in Eqs. \eqref{deltar} and \eqref{Hp}. 
This is well known in the spin-boson model \cite{Leggett1987}. 
Furthermore, using the polaron theory the qubit excitation probability is given by Eq. \eqref{eq:GSExcit}.
Thus, the discussion on $\Delta_r$ directly applies to the existence of excitations in the ground state
because of the coupling to the waveguide.
\\
In Fig. \ref{fig:DeltarPanel} (a) we plot $\Delta_r$
as function of the contact points distance $x$ and the coupling strength $\alpha$ for a giant emitter with $N_c=3$.
Fig. \ref{fig:DeltarPanel} (b) focuses on particular cases and limits of the renormalization of $\Delta_r$.
We have verified that in the limit $x \to 0$ the \emph{dipole approximation} is recovered, i.e. results must reduce to the case $N_c=1$. 
This is a consequence of the normalization used in Eq. \eqref{gk}.
For $N_c=1$, we know that for an Ohmic waveguide $\Delta_r\sim\Delta(\Delta/\omega_c)^{\alpha/(1-\alpha)}$  in the scaling limit $\Delta/\omega_c\ll 1$ \cite{DiazCamacho2016}, which is shown as a dotted line in Fig. \ref{fig:DeltarPanel}(b).
On the other hand, when $x \to \infty$, $\Delta_r$ 
behaves as if the contacts points were \emph{independent}, thus approaching the dipole approximation but with a coupling \emph{per contact} $\alpha \to \alpha/N_c$ (shaded line), showing perfect agreement with the numerical calculation. 
An interesting finding is the appearance of a \emph{distance-dependent localization transition} for a giant emitter, which resembles the one observed for the two-impurity spin boson model \cite{McCutcheon2010, DistantEmitters2021}. This a consequence of the presence of position-dependent couplings in the giant emitter and the competition between the bare qubit energy and dissipation induced by the Ohmic bath. 
%

The polaron calculations predict a more abrupt fall down of $\Delta_r$, in contrast with the single emitter limit(s).
However, it is not
clear that our polaron theory is valid in these (intermediate) ranges of coupling, and the results must be contrasted with other approaches.
Thus, now we resort to a field theoretical argument.
In fact, the existence of a qunatum phase transition in the spin boson model is well studied in the literature \cite{Leggett1987, weiss2012quantum}.
A condition for a symmetry breaking point and thus $\langle \sigma^x \rangle \neq 0$ is that $\int {\rm d} \omega J(\omega)\, \omega^{-2}$ diverges. This happens whenever $J(\omega) \sim \omega^{1-\beta}$ for $0 < \beta < 1$. The Ohmic case ($N_c=1$) lies at the margin \cite{Spohn1985}.  
It is known that, in this case, there is a continuum transition of the Berezinskii–Kosterlitz–Thouless transition type.
This can be proven by mapping the spin-boson model \eqref{spin::boson} to a gas of charges. 
Concretely, the partition function can be approximated by \cite{guinea1998}
\begin{equation}
    Z \sim \exp\left({- 4 \int _0^\beta d\tau_i \int_0 ^\beta  d \tau_j \epsilon_i \epsilon_j  \;  \lambda \left( \frac{\tau_i - \tau_j}{\tau_c} \right)}\right)
\end{equation}
with $\epsilon_i = \pm 1$, $\tau_c = \omega_c^{-1}$  ($\beta^{-1}$ is the temperature) and the \emph{effective} interaction is given by
\begin{equation}
    \frac{ d^2 \lambda(\tau)}{d \tau^2}
    = \omega_c^2 \; \int d \omega J(\omega) \; e^{- \omega \tau},
\end{equation}
which yields $\lambda(\tau) \sim \log(\tau)$ in the dipole approximation for the Ohmic case, thus a Berezinskii–Kosterlitz–Thouless-like transition. 
For a giant atom with arbitrary contact points, the integral on the right side can be computed using the general spectral function in Eq.\eqref{J::omega}, such that
\begin{eqnarray}
    \frac{d^2 \lambda(\tau)}{d\tau^2} \sim \sum_{j,l}^{N_c}\left[\tau+i\left(x_l-x_j\right)/v_{g}\right]^{-2}.
\end{eqnarray}
As an example, for a giant emitter with $N_c=3$ equidistant points it yields,
\begin{equation}
\lambda (\tau) \sim
    -3 \log (\tau )-2 \log \left(\tau ^2+x^2\right)-\log \left(\tau ^2+4 x^2\right),
\end{equation}
i.e., logarithmic interactions persist, one per each leg of the giant emitter. Thus, \emph{confirming the existence of a quantum phase transition in a giant emitter} with arbitrary coupling points.


\subsection{Virtual Photons in the Ground State}

\begin{figure}[!t]
    \centering
    \includegraphics[width=\columnwidth]{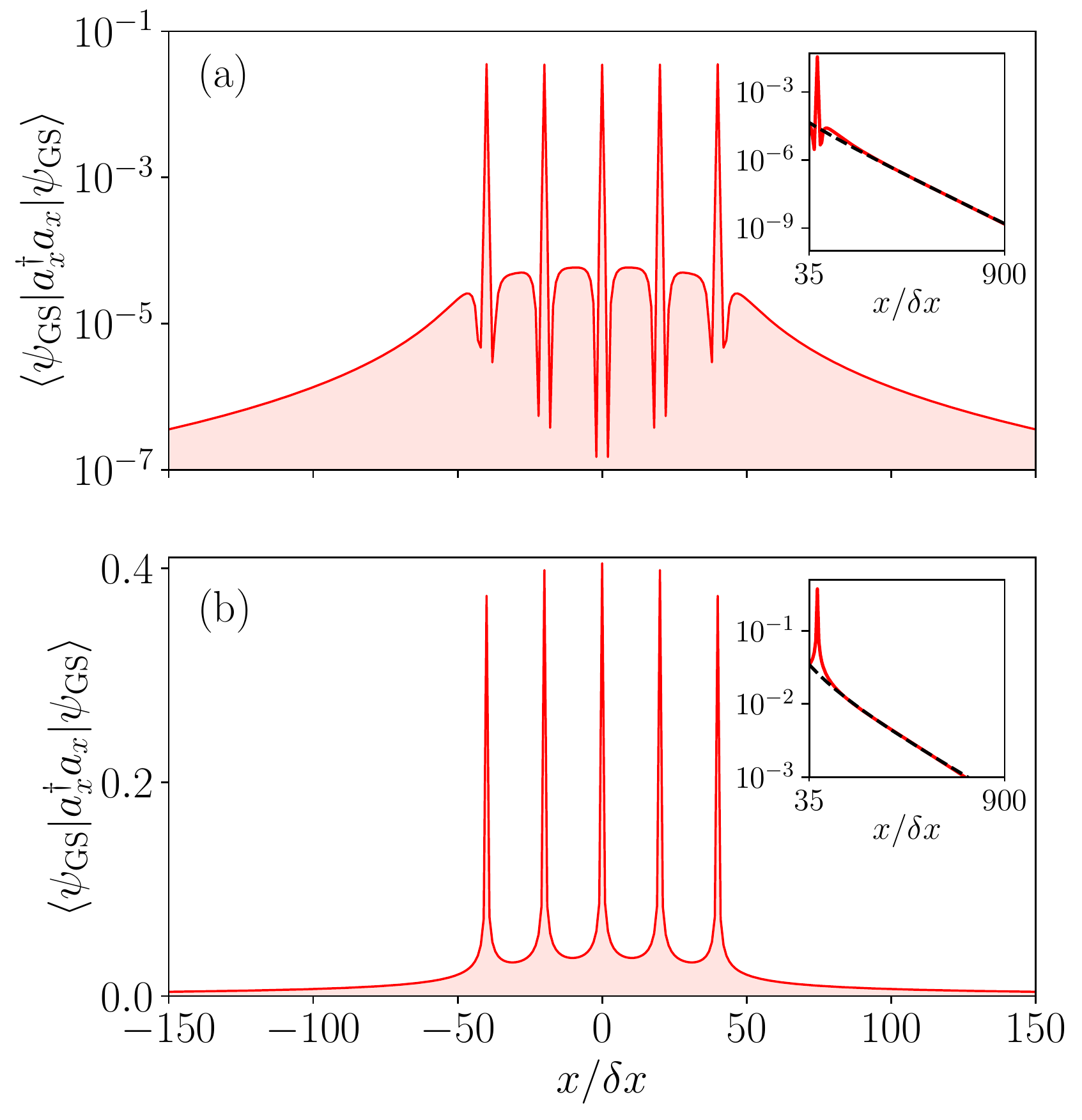}
    \caption{(a) Ground-state photons for a giant atom with five connection points ($N_c = 5$) with coupling strength $\alpha/N_c = 1/5$, in this case $\Delta_r \neq 0$. The inset plot focuses on the profile of the photonic clouds around the right-most coupling point. A fitting into a power-law decay $x^{-a}$, plotted in a black dashed line, leads to the exponent $a \simeq 2.96$. (b) Illustrates the ground-state photons for $\alpha/N_c = 2/5$, corresponding to $\Delta_r = 0$. The decay also fits into a power-law, in this case with $a \simeq 1.09$. For both plots, the distance between the closest couplings is $x = 20 \delta x$, the cutoff frequency is $\omega_c$ = 3, $\Delta = 1$ and the number of modes $N$ = 15001.}
    \label{fig:GSphotons}
\end{figure}

The existence of virtual photons around the contact points at ultrastrong coupling has been hypothesized in \cite{KockumReview2021}. 
Photon localization of the ground state has been successfully studied using polaron and matrix product states simulations for a small emitter in \cite{sanchez2019,JRR2020}, corroborating the usefulness of the variational polaron ansatzes. 
In this section, we describe such photonic clustering for a giant emitter and analyze its spatial profile.

For atoms coupled to cavity-array systems in the USC regime, the photonic cloud generated around the emitter has been found to have an exponentially decaying profile \cite{sanchez2014,sanchez2019,JRR2020}. 
Interestingly, the Ohmic model for the waveguide predicts a power-law decay for the photonic cloud localized around each of the contact points of the giant emitter. 
Furthermore, this power-law decay changes when crossing the QPT.

Using Eq. \eqref{eq:FourierTransf} and at the scaling limit where $\omega_k \approx v_g \lvert k \rvert$, the virtual photons are given by the Fourier transform of $\sqrt{\lvert k \rvert}/(\lvert k \rvert + \Delta_r/v_g)$.
We are interested in the decay of the photonic cloud well away from the connection points, so the corresponding contribution of the integral is that of small-$k$ values.
Therefore, there are two limits of the Fourier transform that interest us. 
Within the delocalized phase $\Delta_r \neq 0$, so we can assume that the contributing $k$ are negligible in front of $\Delta_r/v_g$, leading to a power-law decay with the form $f_x\sim (x-x_j)^{-3/2}$.
Instead, after crossing the quantum phase transition to the localized regime $\Delta_r = 0$ and the decay goes as $f_x \sim (x-x_j)^{-1/2}$.

In Fig. \ref{fig:GSphotons} we plot an example of the ground state photons in real space $\bra{\psi_{\rm GS}}a^{\dagger}_{x}a_{x}\ket{\psi_{\rm GS}} = \lvert f_x \rvert^2$ for both cases, with $N_c = 5$ and $x = 20\delta x$. 
Figure \ref{fig:GSphotons} (a) illustrates the case for $\Delta_r \neq 0$. We observe sharp peaks around each of the coupling points, each of these peaks is surrounded by abrupt dips and a slowly decaying profile. 
The dips can be attributed to the overlap between the sharp peaks and slow decays. 
For this case we predict a power-law decay of the photonic profile away from the emitter scaling as $\sim x^{-3}$. The inset of the figure zooms into the rightmost coupling point and shows a power-law fit in a black shaded line. 
From the fit we recover a decay $\sim x^{-2.96}$ agrees with our prediction.
The other example shown in Fig. \ref{fig:GSphotons} (b) corresponds to $\Delta_r = 0$. 
Here, the peaks become higher and sharper, the dips disappear and the decay becomes slower. 
Again, fitting the profile away from the rightmost coupling point we have a power-law decay $\sim x^{-1.09}$, which perfectly agrees with our analytical estimation.


\section{Relaxation Rate and Lamb Shift} \label{sec:DecayRateLambShift}

\begin{figure}[t]
\centering
\includegraphics[width=\columnwidth]{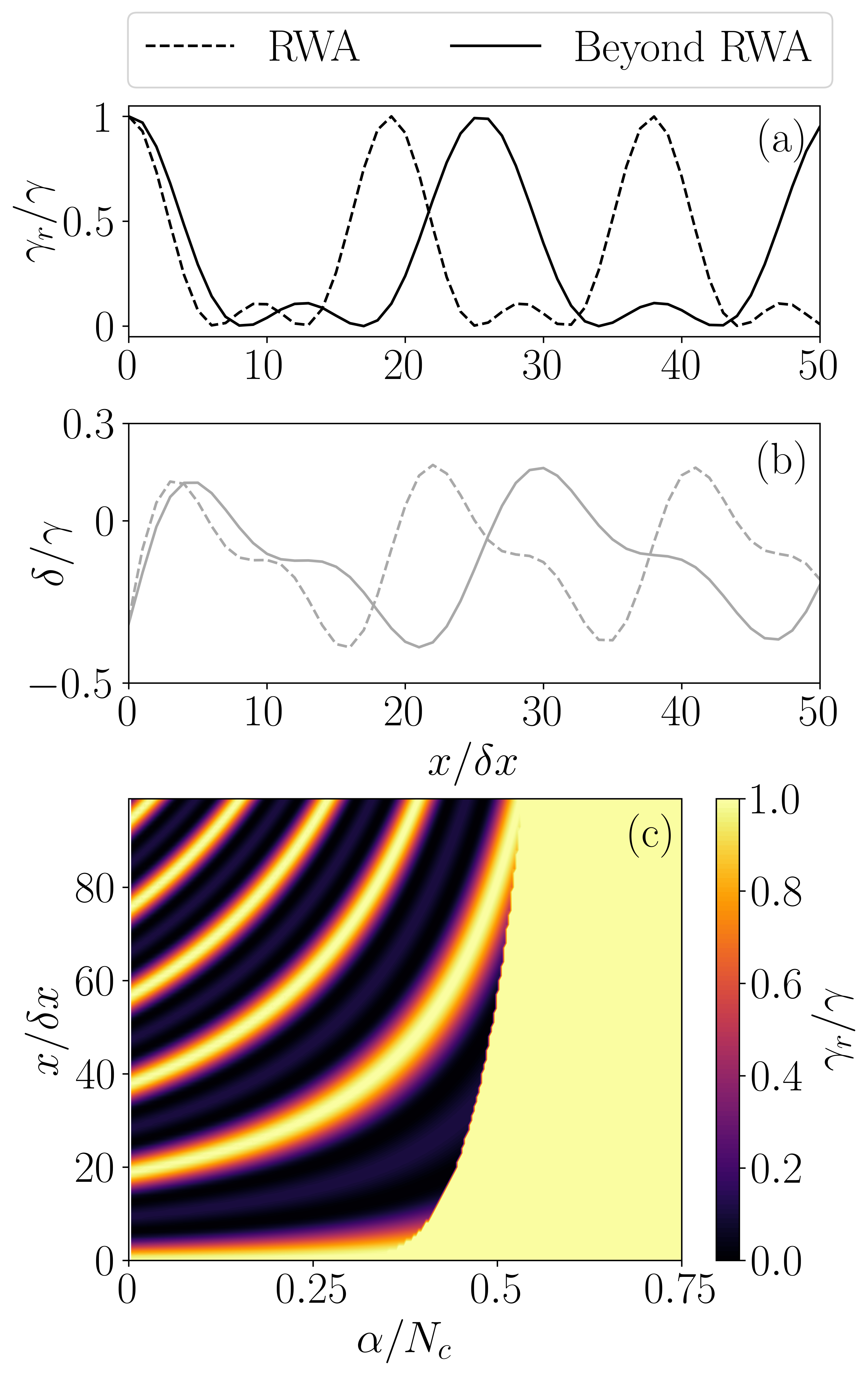}
    \caption{(a) and (b) Scaled effective decay rate and Lamb shift for a giant emitter with three coupling points ($N_c = 3$) and bare qubit frequency $\Delta = 1$ as a function of the spacing between connections. Solid lines indicate the behavior for $\alpha = 0.16$ while dashed lines show a case within the RWA, $\alpha = 0.01$. (c) Full dependence of the effective decay rate of both the coupling strength and the distance between coupling points. 
    \label{fig:RatioGammaUSC}}
\end{figure}

In the simplest approach, the spontaneous emission of an emitter in a continuum is obtained by means of the Fermi's golden rule.
Using second-order perturbation theory (in the light-matter coupling) a two-level system with level splitting $\Delta$ decays with a rate $\gamma = J(\Delta)$. 
Also, the atom frequency is dressed by the Lamb shift $\delta$.

Interestingly, for a giant emitter with multiple contact points, interference effects start to play an important role in the relaxation dynamics.
The fact that the emitter-waveguide interaction is no longer local introduces a new time scale in the system-accounting for the time-delay between different coupling points $\zeta = x/v_g$.
When this time-delay is much smaller than the excited state lifetime of the system as if it had a single coupling point. $\zeta \ll J_{\mathrm{Ohm}}(\Delta)^{-1}$, memory effects can be neglected \cite{KockumPRA2014, KockumReview2021}.
Consequently, an effective relaxation rate $\gamma_r$ and the frequency shift can be obtained in this regime by using the Fermi golden rule, which now depends on the distance between coupling points, and can be engineered to suppress or enhance spontaneous emission. 

In the USC regime, both the emission rates and Lamb shift can be calculated in a similar way as in the perturbative regime.
The only difference is that the formulas are now evaluated at the renormalized frequency $\Delta_r$ instead of the bare one $\Delta$, Cf. Eq. \eqref{deltar} \cite{DiazCamacho2016, Zueco2019}. %
Then,
\begin{eqnarray}
\label{eq:Gammar}
\gamma_r=J(\Delta_r)=J_{\rm{Ohm}}(\Delta_r)G(\Delta_r),
\end{eqnarray}
and
\begin{eqnarray}
\label{eq:LambShift}
\delta=\frac{2\Delta_r^{2}}{\pi}\mathcal{P}\int_{0}^{\infty}d\omega\frac{J(\omega)}{(\Delta_r-\omega)(\omega+\Delta_r)^2}.
\end{eqnarray}

In Fig. \ref{fig:RatioGammaUSC} (a)
we show the normalized relaxation rate as a function of the distance $x/\delta x$ between contact points, for two values of the coupling parameter: $\alpha=0.01$, where we recover the weak coupling or RWA results \cite{KockumPRA2014, KockumReview2021}; and $\alpha=0.16$, where the RWA breaks down.

We observe that increasing the emitter-waveguide coupling beyond RWA produces a \emph{shift in position for the relaxation rate}, displacing characteristic points of destructive and constructive interference. 
This shift is a consequence of the renormalization of the giant emitter frequency and it has to be taken into account in order to observe interference effects in experiments with ultrastrongly coupled giant emitters.

A more complete image of this behavior is given in Fig. \ref{fig:RatioGammaUSC} (b) where the shift is limited by the localization transition appearing at larger values of the coupling (deep strong coupling). 
Therefore, the spontaneous decay in a giant emitter is strongly affected by interference between contact points.
This behavior persists in the USC regime but with values that become strongly modified as the coupling $\alpha$ increases.
Figure \ref{fig:RatioGammaUSC}(c) plots the Lamb shift, reflecting the same shift in position as the relaxation rate in the USC regime. 


\section{Emitter and Field Dynamics}\label{sec:nonMarkDyn}

The effective number-preserving Hamiltonian \eqref{Hp} permits us to work in the single-excitation sector and apply standard RWA methods. 
Using the dynamical polaron \emph{ansatz}, the time-dependent state vector in the polaron frame can be described as \cite{DiazCamacho2016},
\begin{eqnarray}
\ket{\psi^{p}(t)}=c(t)\ket{e}\ket{\mathbf{0}}+\sum_{k}\phi_k(t)\ket{g}a^{\dagger}_{k}\ket{\mathbf{0}}.
\end{eqnarray}
The amplitudes of the polaron state vector satisfy the set of dynamical equations,
\begin{subequations}
\begin{align}
i\dot{\tilde{c}}&=
2\Delta_r\sum_{k}f_{k}\tilde{\phi}_{k}e^{-i(\omega_k-\Delta_r)t},\label{eq:DynSyst1}\\
i\dot{\tilde{\phi}}_{k}&=
2\Delta_r f_{k}\left(\tilde{c}e^{i(\omega_k-\Delta_r)t}+ \sum_l  f_l \tilde{\phi}_l\right), \label{eq:DynSyst2}
\end{align}
\end{subequations}
where we have shifted to different rotating frames $\tilde{c}=e^{i\Delta_r t/2}c$, and $\tilde{\phi}_{k}=e^{i(\omega_k-\Delta_r/2)t}\phi_{k}$, in order to simplify the equations.
Equations \eqref{eq:DynSyst1} and \eqref{eq:DynSyst2} can be integrated numerically, obtaining any observable in the polaron frame. Then, by using relations \eqref{eq:PolLabExcit} or \eqref{eq:PolLabField}, the observables in the Lab frame can be computed.

Before looking at the numerical results, it is convenient to discuss some generalities about the expected dynamical behavior.
For this, we can neglect the contributions of the $V_{\text{local}}$ operator, which only produces a photon frequency shift that does not significantly contribute to the single-excitation dynamics. Besides, it makes further analytical treatment difficult and it is not relevant for the results discussed in this section. 
If this is done, the set of equations are formally equivalent to the one excitation dynamics in RWA models and the Wigner-Weisskopf theory can be directly applied.
Integrating out the photonic degrees of freedom a (non-local) differential equation for $\tilde{c}(t)$ is obtained,
\begin{eqnarray}
\label{eq:EqIntDiff}
\dot{\tilde{c}}=-\frac{2\Delta_r^2}{\pi}\int_{0}^{\infty}\frac{J(\omega)d\omega}{(\omega+\Delta_r)^2}\int_{0}^{t}d\tau\tilde{c}(t-\tau)e^{i(\omega-\Delta_r)\tau}.\nonumber\\
\end{eqnarray}
The dependence of the spectral function on the time delays between coupling points $\zeta$, gives rise to a multiple-time-delay differential equation for the excited state amplitude,
\begin{eqnarray}
\label{eq:EqMovimNM}
\dot{\tilde{c}}(t) &=& - \frac{\gamma}{2 N^2_c}\sum^{N_c-1}_{j=0}(N_c-j)e^{\imag \Delta_r j\zeta} \tilde{c}(t- j \zeta) \Theta(t- j\zeta). \nonumber\\
\end{eqnarray}
Here $\Theta(\cdot)$ is the Heaviside step function. 
The time delays $j \zeta$ introduce new time scales in the system and non-Markovian effects are expected.

\begin{figure}[t]
    \centering
    \includegraphics[width = \columnwidth]{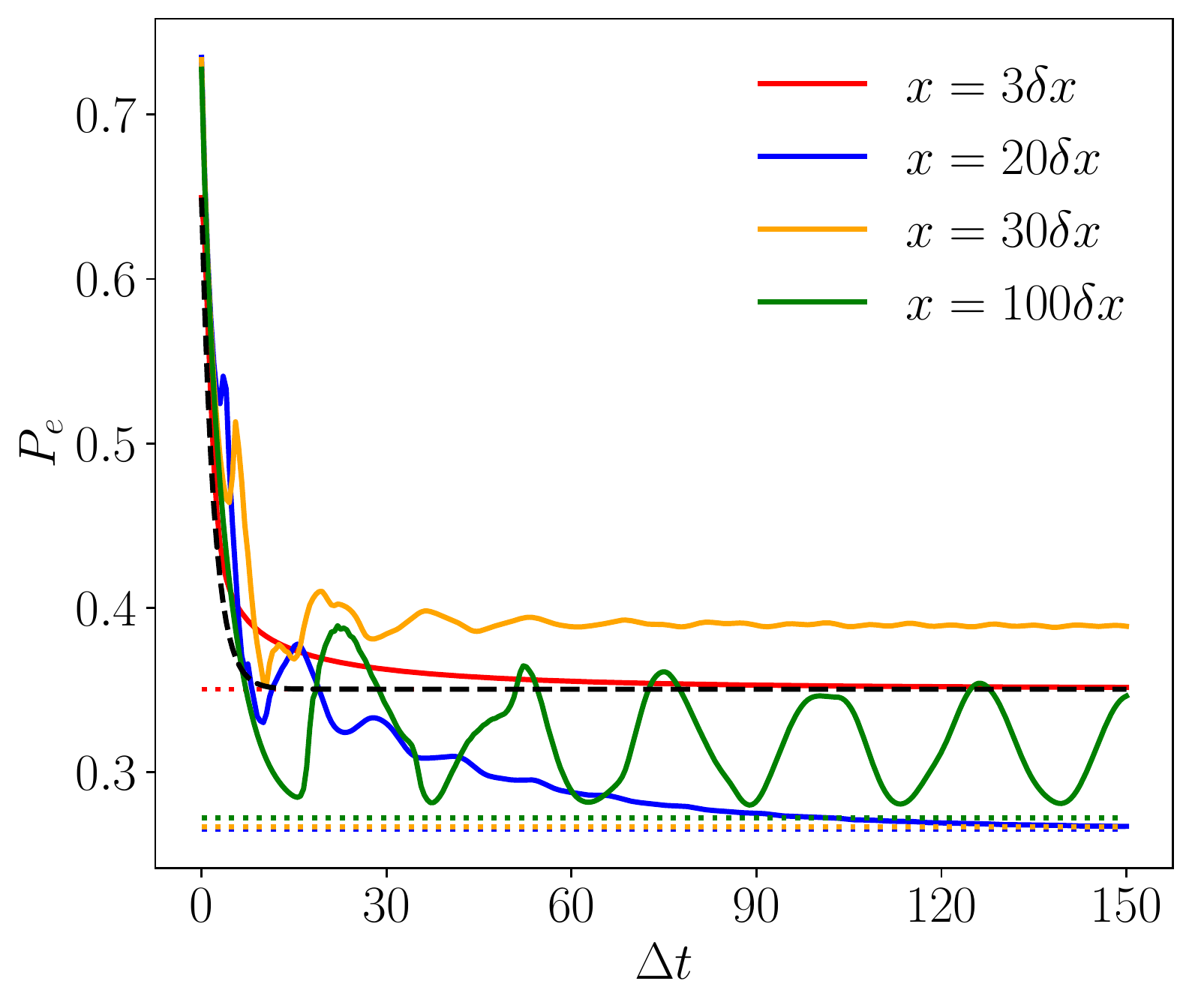}
    \caption{Evolution of the excited state probability of a giant emitter with three coupling points $N_c = 3$, for three different distances between contact points and fixed coupling strength $\alpha = 0.8$, transition energy $\Delta = 1$, cutoff $\omega_c = 6$ and $N = 3000$ modes. The expected equilibrium probabilities $P^{\rm GS}_e$ are illustrated with dotted lines.} 
    \label{fig:decays}
\end{figure}

An analogous time-delay equation was first presented in \cite{KockumPRR2020} within the RWA regime for the same continuous model studied here. 
These type of non-Markovian dynamical equations have also been found in the study of the spontaneous emission in single-end optical fibers \cite{Tuffarelli2013}, atoms in front of reflecting mirrors \cite{Dorner2002}, and two distant emitters in waveguide-QED, within RWA \cite{Kanu2020} and beyond RWA \cite{DistantEmitters2021}. 
In particular, in addition to the relaxation rate previously discussed, oscillations in the emitter dynamics will occur. 
\\
On top of that, already in the RWA regime the existence of bound states for giant atoms has been discussed \cite{KockumPRR2020}.
These can exist even in the absence of band gaps as an interference effect, as seen in Fig. \ref{fig:decays}, due to the spatial separation of coupling points.
Bound states arising from interference effects are also present in the USC regime, as we will show later in numerical simulations. 

Applying a Laplace transform in Eq. \eqref{eq:EqMovimNM} gives us insight on the nature of these bound states.
By defining the excited state amplitude in Laplace space as $\hat{c}(s) = \int^{\infty}_0 dt e^{-st} \tilde{c}(t)$ we have
\begin{equation}
    \label{eq:LaplTransf}
    \hat{c}(s) = \left[s+ \frac{i \Delta_r}{2}+\frac{\gamma}{2N_c^2}\sum^{Nc-1}_{j=0}(N_c-j)e^{(-s+i\Delta_r/2)j|\zeta|}\right]^{-1},
\end{equation}
where we have set $\tilde{c}(0) = 1$ in order to study the spontaneous emission. 
The above dynamical equation in the Laplace space is exactly the same as the obtained for the RWA limit in \cite{KockumPRR2020}, with the difference that the bare qubit frequency $\Delta$ must be replaced by $\Delta_r$.

By definition, bound states do not radiate, thus (if they exist) they are purely imaginary poles of Eq. \eqref{eq:LaplTransf}. 
Searching for purely imaginary poles with the form 
 $s_n = -i2n\pi/(N_c \zeta)$ with $n \in \mathbb{N}$ we obtain
\begin{equation}
\label{eq:DarkStates}
\Delta_r \zeta  = \frac{2n\pi}{N_c} - \frac{J_{\text{Ohm}}(\Delta_r) \zeta }{2N_c}\text{cot}\left(\frac{n \pi}{N_c} \right),
\end{equation}
where $\zeta = x/v_g$ is the time delay between the two closest coupling points.  
It is worth recalling that all these relations neglect the local $V_{\rm local}$ operator, Cf. Hamiltonian \eqref{Hp}.
They are, however, a good estimation for understanding the emitter dynamics and locating the appearance of bound states in the parameter space of the model. 
In particular, that their existence requires a finite time delay ($\zeta$) and that the renormalized atom frequency and spontaneous emission play a role.

Both the existence of bound states and non-Markovian dynamics in the USC regime can be proven by monitoring the spontaneous emission.
In doing so we assume the atom-waveguide at the GS,
then the qubit is driven within a $\pi$-pulse. After the pulse the wavefunction is given by $\ket{\psi(0)} =\sigma^{+}\ket{\rm GS}$. 
Since $[\sigma^{x}, U_p] = 0$, we may work in the single-excitation manifold in the polaron picture. 
Therefore, 
we can numerically integrate \eqref{eq:DynSyst1} and \eqref{eq:DynSyst2}, including the $V_{\text{local}}$ terms, and transform back to the Lab frame using \eqref{eq:PolLabExcit}. 
In figure \ref{fig:decays} we plot the spontaneous emission. 
We notice that $P_e(t=0) \neq 1$, since for our initial state $P_e= 1/2(1+ \Delta_r/\Delta)$, Cf. Eq. \eqref{eq:PolLabExcit}. %
Furthermore,
each of the plotted decay processes has a different equilibrium excited state occupancy, as given by Eq. \eqref{eq:GSExcit}.

\begin{figure*}[!t]
    \centering
    \includegraphics[width = \linewidth, keepaspectratio]{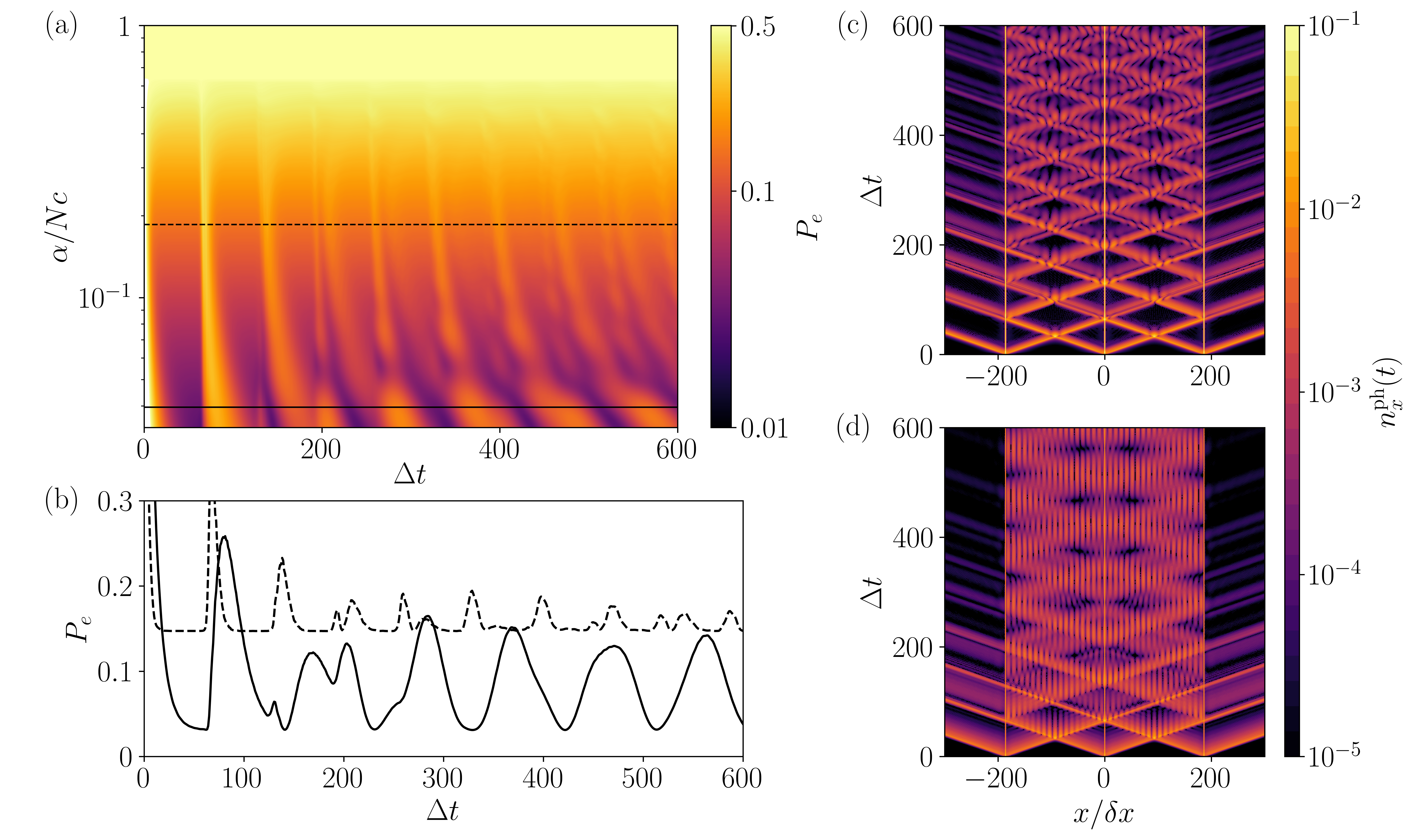}
    \caption{(a) Time evolution of the excited state population $P_e$ with increasing coupling $\alpha$. Black horizontal lines point to the two cases highlighted in the other plots of the figure. 
    (b) Plot of the spontaneous emission into an oscillating bound state in the RWA for $\alpha \simeq 0.118$ (solid line) and for an increased coupling of $\alpha \simeq 0.557$, within USC (dashed line). The field emitted into the waveguide for these cases is plot in (d) and (c) respectively.
    These simulations are carried out for a giant atom with three connections $N_c = 3$, with the parameters $\zeta = 186\delta x$, $\omega_c = 3$, $\Delta = 1$ and number of modes $N = 3002$.}
    \label{fig:OBSPanel}
\end{figure*}
As shown in Fig. \ref{fig:decays} for $x = 3\delta x$, 
short time delays between coupling points lead to relaxations that can be closely described by an exponential decay
defined by the effective spontaneous emission rate in Eq. \eqref{eq:Gammar}.
The evolution of the excited state occupancy can be well approximated by:
\begin{equation}
    \label{eq:PolaronDecay}
    P_e \approx P^{\rm GS}_e + \frac{\Delta_r}{\Delta}e^{-\gamma_r t},
\end{equation}
plotted in the dashed black line in Fig. \ref{fig:decays}.
The lack of major non-Markovian effects can be seen via Eq. \eqref{eq:LaplTransf}, as the limit ${\zeta \to 0}$ leads to a time evolution governed by the decay rate $\gamma_r$. 

For a sufficiently large distance between coupling points, non-Markovianity takes a central role, as illustrated by the decay for $x = 20 \delta x$ in Fig. \ref{fig:decays}. 
Initially, it has an approximately exponential decay given by $J_\mathrm{Ohm}(\Delta_r)$, until the emitted light reaches a coupling point.
Then, an oscillatory behavior rises as the light emitted from one connection point is partially reabsorbed and emitted back to the waveguide by another contact point.
At longer times, these oscillations become damped as the energy is gradually emitted outside the atom, until the system reaches the equilibrium at the corresponding $P^{\rm GS}_e$.

We encounter a different behavior for $x = 30\delta x$ in Fig. \ref{fig:decays}. Due to the interference of the field emitted from each coupling point, the system relaxes to a bound state, as signaled by the difference in occupancy from the ground state once the evolution reaches the equilibrium.
This comes as a direct consequence of the initial excited state having a non-zero overlap with bound states for these parameters.

Furthermore, Fig. \ref{fig:decays} illustrates yet a another type of decay.
For $x = 100 \delta x$ we find long-lived oscillations around an equilibrium value higher than $P^{\rm GS}_{e}$.
This is reminiscent of the reported oscillating bound states in the RWA \cite{KockumPRR2020}.
In the next section we discuss how these oscillating bound states behave whenever the coupling cannot be treated perturbatively.


\subsection{Oscillating bound states} 

Some interest has been aroused in the existence of \emph{oscillating} bound states \cite{KockumPRR2020, noachtar2022nonperturbative}.
They originate from the interplay of two coexisting bound states.
Consequently, part of the field emitted during the spontaneous emission process is trapped while oscillating between the coupling points of the emitter.
In USC, approximate oscillating bound states are found by searching two coexisting solutions of Eq. \eqref{eq:DarkStates}.
Due to the dependence of $\Delta_r$ with $\alpha$ and $\zeta$, via the variational parameters $f_k$, the existence of two solutions for the same set of parameters cannot be analytically proven.
This numerically requires fine tuning, which at most allows for the prediction of oscillating bound states with large but finite lifetime in USC, as illustrated in Fig. \ref{fig:decays} for $x = 100\delta x$. 

Figure \ref{fig:OBSPanel} (a) illustrates the excited state population $P_e$ of a giant atom with increasing coupling strength $\alpha$. The rest of the parameters are set so that in the RWA regime the excited state decays into an oscillating bound state. 
It is clear that the oscillating bound state found for the lower couplings is lost as $\alpha$ increases. 
Figure \ref{fig:OBSPanel} (b) focuses on two specific values of the coupling strength showing the long time behavior of the oscillating bound state and its counterpart at higher coupling, for which the population revivals slowly decrease in amplitude. 
The field evolutions corresponding to both cases are plotted in Figs. \ref{fig:OBSPanel} (c) and (d), respectively.

Within the RWA, fixing the distance between coupling points and increasing the coupling should eventually reach another pair of simultaneous solutions for Eq. \eqref{eq:DarkStates} and therefore an oscillating bound state \cite{KockumPRR2020}.
In contrast, our simulations indicate that entering the USC regime leads to the loss of the oscillating bound state due to the renormalization of $\Delta_r$ and eventually to the quantum phase transition. 
This same trend has also been recently reported in \cite{noachtar2022nonperturbative} where matrix product states simulations showed this same breakdown of the periodicity for the existence of oscillating bound states in parameter space with the coupling strength.

%
\section{Summary and Conclusions}\label{sec:Conclusion}

In this work we have developed a semianalytical approach for the low energy sector of giant emitters in the ultrastrong regime based on polaron-like methods. 
In particular, we have focused on a single giant atom coupled via $N_c$ connection points to an Ohmic waveguide.

We have characterized the ground state of the system.  
In particular we have analyzed the virtual photons surrounding each of the coupling points.
The latter decays spatially away from the connection points to the waveguide as a power law, unlike what occurs for point-like emitters in a cavity array \cite{JRR2020}.
We also have studied the renormalization of the atomic frequency $\Delta_r$ and shown that the system exhibits a localization-delocalization quantum phase transition which is dependent not only on the coupling strength $\alpha$ but also on the distance between coupling points.

For the dynamics of the system we have focused on the spontaneous emission.
We have derived analytical expressions for the Lamb-shift $\delta$ and effective decay rate $\gamma_r$ which characterize the early evolution of the system whenever its lifetime is much larger than the distance between coupling points.
Both of these values were reported to have a periodic behavior with the distance between coupling points in the RWA regime \cite{KockumPRA2014}.
We find that this periodicity is lost in USC due to the renormalization of $\Delta_r$.

We were able to fully characterize the dynamics within the polaron frame, providing an approximate analytical expression for the evolution of the excited state amplitude.
We find that some of the non-Markovian dynamics found in the RWA, such as the non-exponential decay \cite{KockumPRA2017} and bound states rising from the interference of the spontaneous emission from different coupling points \cite{KockumPRR2020}, still hold in the USC regime.
However, other behaviors such as the recurrence of oscillating bound states as the coupling increases \cite{KockumPRR2020} are lost when entering the USC regime. 
Instead, as the localization-delocalization transition is approached by increasing the coupling strength $\alpha$, the oscillations in the excited state occupancy have a sharp drop in amplitude, becoming irregular in time and eventually disappear.

\begin{acknowledgments}
The authors acknowledge funding from the EU (COST Action 15128 MOLSPIN, QUANTERA SUMO and FET-OPEN Grant 862893 FATMOLS), the Spanish Government Grants MAT2017-88358-C3-1-R and PID2020-115221GB-C41/AEI/10.13039/501100011033, the Gobierno de Arag\'on (Grant E09-17R Q-MAD) and the CSIC Quantum Technologies Platform PTI-001.
C.A.G.-G. acknowledges funding
from the program Acciones de Dinamización
“Europa Excelencia” Grant No. EUR2019-103823.
F.N. is supported in part by:
Nippon Telegraph and Telephone Corporation (NTT) Research,
the Japan Science and Technology Agency (JST) [via
the Quantum Leap Flagship Program (Q-LEAP), and
the Moonshot R\&D Grant Number JPMJMS2061],
the Japan Society for the Promotion of Science (JSPS)
[via the Grants-in-Aid for Scientific Research (KAKENHI) Grant No. JP20H00134],
the Army Research Office (ARO) (Grant No. W911NF-18-1-0358),
the Asian Office of Aerospace Research and Development (AOARD) (via Grant No. FA2386-20-1-4069), and
the Foundational Questions Institute Fund (FQXi) via Grant No. FQXi-IAF19-06.
\end{acknowledgments}


\bibliography{ref}
\end{document}